\documentstyle[twoside,fleqn,espcrc2,epsfig]{article}

\def\slash#1{\mbox{$\not \!\! #1$}}

\def\ct#1{\mbox{\cal #1}}
\def\spose#1{\hbox to 0pt{#1\hss}}
\def\ltapprox{\mathrel{\spose{\lower 3pt\hbox{$\mathchar"218$}}
 \raise 2.0pt\hbox{$\mathchar"13C$}}}
\def\gtapprox{\mathrel{\spose{\lower 3pt\hbox{$\mathchar"218$}}
 \raise 2.0pt\hbox{$\mathchar"13E$}}}
\def\inapprox{\mathrel{\spose{\lower 3pt\hbox{$\mathchar"218$}}
 \raise 2.0pt\hbox{$\mathchar"232$}}}

\renewcommand{\Pr}{\hat{\mbox{I}\!\!\mbox{P}}}

\newcommand{\Dim}{\mbox{dim}\,}
\newcommand{\Tr}{\mbox{Tr}\,}

\newcommand{\sub}{\mbox{\scriptsize sub}}
\newcommand{\tree}{\mbox{\scriptsize tree}}
\newcommand{\true}{\mbox{\scriptsize true}}

\newcommand{\RI}{\mbox{\scriptsize RI}}
\newcommand{\QCD}{\mbox{\scriptsize QCD}}

\newcommand{\latt}{\mbox{\small latt}}
\newcommand{\phys}{\mbox{\small phys}}

\newcommand{\etal}{{\em et al.}}

\newcommand{\CMP}{Comm. Math. Phys.}
\newcommand{\NPB}{Nucl. Phys. B}
\newcommand{\PLB}{Phys. Lett. B}
\newcommand{\PRD}{Phys. Rev. D}
\newcommand{\ZPC}{Z. Phys. C}

\newcommand{\nn}{\nonumber}
\newcommand{\<}{\langle}
\renewcommand{\>}{\rangle}

\newcommand{\ctZ}{{\cal Z}}
\newcommand{\be}{\begin{equation}}
\newcommand{\ee}{\end{equation}}
\newcommand{\bea}{\begin{eqnarray}}
\newcommand{\eea}{\end{eqnarray}}

\title{Status and Perspectives of Non-perturbative Renormalization in 
Weak Decays.}

\author{
M.~Talevi\address{Department of Physics \&\ Astronomy, University of Edinburgh,
The King's Buildings, EH9 3JZ (UK)}
\thanks{Invited talk at the International Workshop 
``Lattice QCD on Parallel Computers'', 10-15 March 1997, Tsukuba (Japan)}
}

\begin{document}

\begin{abstract}
We discuss the status and the problems related to the application of the 
off-shell non-perturbative renormalization method in a fixed gauge to operators
relevant to weak decays.  In particular, we critically reappraise the method 
recently proposed for the $\Delta I=1/2$ rule.
We also present a general analysis of the renormalization for the 
$\Delta I=3/2$ operators, and apply it to the $\Delta S=2$ operator.
\end{abstract}

\maketitle

\section{Introduction}

Lattice QCD is a unique, systematically improvable, method for computing 
matrix elements from first principles, and has proven a powerful and appealing
approach.  In spite of the successes, progress has been slow due to the 
presence of systematic effects, such as discretization and 
higher-order renormalization effects.  
The Symanzik improvement program \cite{Symanzik,LW} 
is an attrative method which allows to reduce discretization order by order in
$a$ in physical quantities.  The improvement coefficients were computed 
at first in perturbation theory (PT) at lowest order,
reducing the error from $O(a)$ to $O(g_0^2a)$ \cite{SW,Heatlie},
and recently non-perturbatively, achieving a full $O(a^2)$ improvement
\cite{Luscher}.  In parallel, there has been significant progess in the 
development of non-perturbative (NP) methods \cite{Martinelli,Jansen,JLQCD}.
It is by now generally accepted that NP methods in the 
renormalization of lattice operators yield reliable and accurate results
and should be used whenever possible.  In the following, we concentrate on the 
applications of the non-perturbative renormalization method of 
ref.~\cite{Martinelli}, hereafter refered to as NPM. 
The other methods are discussed by Sommer \cite{Sommer} and Kuramashi 
\cite{Kuramashi} in this Workshop.

Renormalization of lattice operators is a crucial ingredient in the calculation
of physical weak matrix elements on the lattice.
A physical amplitude $A_{\alpha \rightarrow \beta}$ of a weak transition
$\alpha \rightarrow \beta$ is calculated via the Operator Product Expansion 
(OPE) by
\begin{equation}
A_{\alpha \rightarrow \beta} = C_W(M_W/\mu) \< \alpha | \hat O(\mu) |\beta \>
\label{eq:ope}
\end{equation}
where $C_W$ is the Wilson coefficient of the OPE, $M_W$ is the mass of the
$W$ boson, 
$\mu$ is the renormalization scale and $\<\alpha|\hat O(\mu)|\beta \>$ is 
the matrix element of the renormalized operator (at the scale $\mu$) relevant 
to the physical process.
The Wilson coefficient $C_W(M_W/\mu)$ contains the short-distance information 
and can be calculated in PT in the continuum at the 
renormalization scale $\mu$. 
The matrix element contains the long-distance dynamics and thus must 
be calculated non-perturbatively on the lattice.
Renormalization relates the regularized lattice matrix elements 
to its continuum counterpart.  

On the lattice, chiral symmetry is explicitly broken with Wilson-like fermions.
The possibility of recovering the chiral symmetry in the continuum limit 
was shown in \cite{Bochicchio}.  The general prescription is to subtract 
from the bare operator $O(a)$ all the operators of dimension less or equal
than $O(a)$, which have the same quantum numbers conserved by the 
regularization,
\begin{eqnarray}
&&\< \alpha | \hat O(\mu) | \beta \> = \nn \\
&&\lim_{a\to 0} \< \alpha | Z_O(\mu a)[O(a)+\sum_i Z_i O_i(a)] |\beta \>,
\end{eqnarray}
where the subtracted operators $O_i$ are not constrained to the same chiral
representation of $O(a)$. If these operators have lower dimension than $O$, 
the mixing constants are power-divergent in the 
cutoff, $Z_i\sim 1/a^d$ with $d=\Dim[O]-\Dim[O_i]>0$.   
These divergent factors can pick up
exponentially small contributions in the coupling $\alpha_s$, yielding a finite
contribution as $a\to 0$, i.e. 
\begin{equation}
\frac{1}{a} e^{-1/\alpha_s(a)}\sim \Lambda_{\QCD}.
\end{equation} 
These divergences must be subtracted in a completely NP way.

\section{Operators and phenomenology}

There are a number of four-fermion operators which are relevant to
different physical processes of phenomenological interest.  They all
have in common that their renormalization suffers from the mixing with
chiral violating form factors induced by the Wilson term.  We can divide
them in three broad classes:
\begin{itemize}
\item  LL operators: the $\Delta I=3/2$ components are necessary for
the $B_K$ parameter, which enters in the study of CP violation
in $K^0--\bar K^0$ mixing.  It is important to obtain a precise and
reliable result, as with the measured value of the top quark
mass, it enables us to limit
the range of values of the CP-violation phase $\delta$.

The $\Delta I=1/2$ components on the other hand are relevant to the
study of the octect enhancement in $K \rightarrow \pi \pi$ decay, 
whose quantitative understanding still defies theorists.
The difference with the $\Delta I=3/2$ case is that the bare operator
are allowed to mix with operators of lower 
dimensionality and hence with coefficients that are power-divergent 
\cite{Maiani,DI=1/2}.  

\item LR operators:  the $I=3/2$ part of the LR operators which appear in the
effective weak Hamiltonian due to the electromagnetic penguin diagrams 
are the only operators which give rise to an imaginary part in the
$K^+\to \pi^+\pi^0$ amplitude, thus yielding the dominant contribution to 
$\epsilon'/\epsilon$.  This contribution is usually expressed by the
$B_7$ and $B_8$ parameters.

The $I=1/2$ part of these operators have penguin contractions which make
them too hard to handle at present.  Note also that in presence of the heavy
top quark, there is no GIM suppression.

\item $\Delta B=2$ operators: these are the chiral partners of the 
$\Delta I=3/2$, and have recently been proposed in the study of 
flavour-changing neutral currents in the supersymmetric extensions
of the Standard Model \cite{Babar}.
\end{itemize}

\section{The method}

In the NPM, the renormalization conditions are applied directly to the 
Green functions of quarks and gluons, in a fixed gauge, with given off-shell
external states of large virtualities \cite{Martinelli}.  
The method mimicks what is usually
done in the perturbative calculation, but the Green functions are evaluated
in a NP fashion from Monte Carlo simulations.

To give the flavour of the method, let us consider the simplified case of a 
multiplicatively renormalizable operator, 
e.g. a two-quark operator $O=\bar q \Gamma q$.
Given the bare lattice operator $O^{\latt}(a)$, the renormalization condition
we impose is \cite{Martinelli} 
\begin{eqnarray}
&&Z^{\latt}_{\RI}(\mu a)\<p|O^{\latt}(a)|p\>|_{p^2=\mu^2} \nn \\
&&=\<p|O^{\latt}(a)|p\>|_{p^2=\mu^2}^{\tree},
\label{eq:Z_RI(mu)}
\end{eqnarray}
where $\<p|\cdots|p\>$ denotes the matrix element of external quarks of momenta
$p$ which can be calculated non-perturbatively in the QCD coupling via 
Monte Carlo simulations \cite{Martinelli}.  
The renormalized operator obtained with the NPM is then
\begin{equation}
\hat O_{\RI}(\mu)=Z^{\latt}_{\RI}(\mu a)O^{\latt}(a),
\label{eq:O_RI(mu)}
\end{equation}
which depends on the external states and the gauge, but not on method used to 
regulate the ultra-violet divergences.  
To stress this point, we call the NP renormalization scheme 
Regularization Independent (RI) \cite{eps'/eps}.  The physical operator 
\begin{equation}
O^{\phys}(M_W)=C_{\RI}(M_W/\mu)\hat O_{\RI}(\mu)
\label{eq:O^phys}
\end{equation}
is independent of external momenta and gauge 
(up to higher orders in continuum PT and lattice systematic effects) if the
Wilson coefficient function $C_{\RI}(M_W/\mu)$ in the RI scheme is calculated 
with the same external momenta and gauge of $\hat O_{\RI}(\mu)$. 
The advantage of the RI scheme is that it completely avoids the use of 
lattice PT, which is expected to have a worse convergence than the continuum
expansion \cite{Lepage}.  The coefficient function 
$C_{\RI}(M_W/\mu)$ are instead calculated in continuum PT, 
which cannot be avoided since the Wilson OPE is defined perturbatively.  

The phenomenologically more interesting case of four-fermion operators
are in general not multiplicatively renormalizable.
The operators which need to be subtracted
are dictated by the symmetries of the action:  charge conjugation (C),
parity (P) and $s\leftrightarrow d$ flavour switching 
symmmetry (S) \cite{Bernard}.

The main advantage of the NPM is its generality, being valid for any 
composite operator, as long as we can can find ({\em a posteriori}) 
a window in the range of renormalization scales $\mu$ such that 
$\Lambda_{\QCD}\ll\mu\ll O(1/a)$, in order to keep under control both the
higher-order effects in the (continuum) perturbative calculation of 
$C_{\RI}$ and discretization errors \cite{Martinelli}.  
We stress that this requirement is common to all NP methods on the lattice 
which work at a single value of the lattice spacing.  The alternative would
be matching from one value of the coupling to another, as done for example in
the method of ref.~\cite{Jansen}.  The main disadvantage of the NPM is the
necessity of gauge-fixing, which leaves a residual less constraining symmetry
to dictate the form of the mixing, i.e.\ BRST-invariance.

\section{$\Delta I=1/2$: Seeking new ideas}

A quantitative theoretical understanding of the $\Delta I=1/2$ rule in 
$K\to\pi\pi$ decays has proven to be a formidable task 
since the calculation of hadronic matrix 
elements in the low-energy NP regime is needed.
Let us review the strategies proposed so far \cite{Maiani,DI=1/2}.

In the continuum, with an active charm quark and the GIM mechanism at work,
the operator basis given by
\begin{equation}
O^{\pm}_{LL}= \frac{1}{2}[(\bar s d)_L(\bar u u)_L
                      \pm (\bar s u)_L(\bar u d)_L] - (u\to c).
\label{eq:OpmGIM}
\end{equation}
In the framework of lattice QCD with Wilson-like fermions,
the renormalization strategy is complicated by chiral symmetry breaking.
In fact, the Wilson term induces the mixing of $O^{\pm}_{LL}$ with 
lower-dimensional operators,
with power-divergent coefficients, which need to be subtracted 
non-perturbatively. 
In the following, we shall concentrate only on the octet component of 
$O^{\pm}_{LL}$, which we will denote with $O^{\pm}_0$.  The lattice 
penguin operators, being 
proportional to $(m_c^2-m_u^2)a^2\ll 1$, will be neglected in the following. 

We are now faced with the problem of calculating the four-point matrix elements
$\<\pi^+\pi^-|O^{\pm}_0|K^0\>$.  The stardard approach is to rely on lowest 
order chiral PT to relate them to the more tractable 
three-point and two-point matrix elements \cite{Maiani}:
\begin{equation}
\<\pi^+\pi^-|O^{\pm}_0|K^0\>=i\gamma^{(\pm)}(m_K^2-m_\pi^2)/f_K 
\end{equation}
where $\gamma^{(\pm)}$ are obtained from 
\[
\left\{
\begin{array}{l}
\<\pi^+(p)|O^{\pm}_0|K^0(q)\>=\gamma^{(\pm)}(p\cdot q) 
- \delta^{(\pm)} m_\pi^2/f_K \\
\<0|O^{\pm}_0|K^0\>=i\delta^{(\pm)}(m_K^2-m_\pi^2) 
\end{array}
\right.
\]
In this approach, one relies on the calculation the $K\to \pi$ matrix element,
which only picks up a contribution from the parity-conserving (PC) part
of the operators.  Exploiting CPS symmetry, we obtain in the PC
sector a renormalization structure of the form \cite{DI=1/2}
\begin{equation}
\widehat O^{\pm}_{PC}=
Z^{\pm}_{PC}\left[O^{\pm}_0+\sum_{i=1}^4 Z^{\pm}_iO^{\pm}_i
+Z^{\pm}_\sigma O_\sigma+Z^{\pm}_SO_S\right]
\label{eq:hatOpm}
\end{equation}
where $O^{\pm}_0$ are the PC bare operators,
$O^{\pm}_i,\ i=1,\ldots,4$ are dimension-six operators of wrong 
chirality (cf.~sec.~\ref{sec:d=6}), $O_\sigma$ is the magnetic operator 
$\bar s\sigma_{\mu\nu}F_{\mu\nu}d$ and $O_S$ is a dimension-three scalar 
density $\bar s d$.  By GIM and power-counting, 
$Z^{\pm}_\sigma\propto (m_c-m_u)$ and $Z^{\pm}_S\propto (m_c-m_u)/a^2$.
Thus, while the coefficient of the magnetic operator can in principle be 
calculated in PT, though it involves a two-loop calculation 
and is very complicated \cite{Curci}, the coefficient of the scalar
density is power-divergent and can only be reliably calculated in a NP
fashion.

There are in principle several NP approaches for calculating the mixing 
coefficients:
\begin{enumerate}
\item by imposing the Ward Identities on physical hadronic states 
\cite{Maiani};
\item by imposing the Ward Identities on quark states, as done for the 
$\Delta S=2$ operator in \cite{JLQCD};
\item by the NPM \cite{DI=1/2}.
\end{enumerate}
 
The principal drawback of the method 1.\ is that it looses predictive power
as the number of coefficients to determine gets large, as in the present case.
Methods 2.\ and 3.\ are equivalent in the region of large $\mu$ 
(cf. sec.~\ref{sec:d=6}), and both require the inclusion of 
\begin{itemize}
\item operators that vanish by the equations of motion, because the 
renormalization conditions are imposed on off-shell Green functions.
\end{itemize}
Some examples are
\begin{equation}
\bar s(\slash{D}+m_d)d, \ \bar s(\slash{D}+m_d)^2d, 
\end{equation}
which, by GIM, generate counterterms proportional to $(m_c-m_u)/a$ and 
$(m_c-m_u)$, respectively.  
When inserted in correlations, these terms give a finite 
contribution to the subtraction coefficient $C_S$ of the scalar density 
$\bar sd$. In PT, this problem is of course also present, 
but the spurious contributions can be eliminated by looking at the momentum 
dependence
\[
C_S=C_S^{\true}+f+g(\slash{p}+\slash{p}')+h(p^2+p^{\prime 2}+bp\cdot p'),
\]
where $f,g,h$ are calculable functions of $m_s$ and $m_d$ and $p,p'$ are the
momenta of the two external legs.  In principle, this could also be attempted
in a NP approach, but it highly unlikely to be able to achieve the necessary
accuracy. 
\begin{itemize}
\item operators are not gauge invariant, because the off-shell Green functions 
are calculated in a fixed gauge.  
\end{itemize}
The non-gauge invariant operators that may mix are dictated by lattice 
BRST symmetry \cite{LesHouches}.
Some of the these operators have been classified in ref.~\cite{Ross}, e.g.
\begin{equation}
\bar s\slash{\partial}\slash{A}(\slash{D}+m_d)d, \ 
\bar s(\slash{D}+m_d)\slash{\partial}\slash{A}d. 
\end{equation}
Again all of these operators need to be taken into account as they 
give a finite contribution to the subtraction coefficient.

The application of the NPM without these additional operators has been 
outlined in ref.~\cite{DI=1/2}.  
Let us recall it to give a flavour of its complexity.
According to the NPM, the mixing $Z$'s are determined by finding a set of 
projectors on the tree-level amputated Green functions (GF), with off-shell
quark and gluon external states,
the choice of which depends on the nature of the operators at hand.
For the $\Delta I=1/2$ operators we choose the following set of external 
states: $q\bar q$, $q\bar qg$, $q\bar qq\bar q$, with the momenta given 
below in eq.~(\ref{eq:6-Z-system}).  
For each choice of external states, i.e. for each different set of GF,
we need different type of projectors.  Let us denote with $\Pr_S$ 
the projector on the $q\bar q$ GF of the operator $O_S$, 
with $\Pr_\sigma$ the projector on the $q\bar qg$ GF of the 
operator $O_\sigma$, and with $\Pr^{\pm}_j,\ j=1,\ldots,4$ the set of mutually
orthogonal projectors on the operators $O_i,\ i=1,\ldots,4$ \cite{DI=3/2}.
Applying the projectors to the corresponding NP GF
of the renormalized operators $\widehat O^{\pm}$,
with an appropriate choice of the external states, 
we require that the renormalized operators be proportional to the bare 
operators, $\widehat O^{\pm}(\mu)\propto O^{\pm}_0(a)$ (up to terms of 
${\cal O}(a)$), i.e.\ we impose the following renormalization conditions
(trace over colour and spin is understood in the projection operation):
\begin{equation}
\begin{array}{l}
\Pr_S\<q(p)|\widehat O^{\pm}|\bar q(p)\>=0  \\
\Pr_\sigma\<q(p-k)g(k)|\widehat O^{\pm}|\bar q(p)\>=0  \\
\Pr^{\pm}_j\<q(p)\bar q(p)|\widehat O^{\pm}|q(p)\bar q(p)\>=0,\ j=1,\ldots,4  
\end{array}
\label{eq:6-Z-system}
\end{equation}
where $p$ and $k$ denote the momentum of the external quark and gluon legs.
The system of equations (\ref{eq:6-Z-system}) in principle completely 
determines in a NP way the renormalization constants, 
as we have six conditions 
(non-homogeneous due to the matrix elements of $O^{\pm}_0$, 
cf.~eq.~(\ref{eq:hatOpm})) in six unknown mixing constants, 
$Z^{\pm}_i,\ i=1,\ldots,4,Z^{\pm}_\sigma,Z^{\pm}_S$.

Unfortunately, since solving eq.~(\ref{eq:6-Z-system}) involves
delicate cancellations between large contributions, it results
in a very noisy determination, even with large statistics.
The main computational difficulty lies in the calculation of the GF
with penguin contractions.
The need to include the operators which vanish by the equations of motions
and are not gauge invariant renders the application of the NPM, which was 
already complicated without them, highly impractical.   We conclude that
the stardard positive parity methods are not a viable way of approaching
the $\Delta I=1/2$ rule and new ideas are needed \cite{Bernard,DI=1/2New}.

\section{$\Delta I=3/2$: A general analysis of dimension-six mixing}
\label{sec:d=6}

We now turn to discuss the renormalization of $\Delta I=3/2$ operators.
They differ from the $\Delta I=1/2$ operators in that there are no 
lower-dimensional operators with the same flavour content with which they
can mix.
This implies that we need not take into consideration the operators
that vanish by the equations of motion or the non-gauge invariant ones
as they can only affect the mixing with the lower dimensional operators.

In order to address this problem, it is convenient to work
with 4 distinct fermion flavours $\psi_f,\ f=1,\dots,4$, 
of degenerate mass. Once the mixing of the dimension-six
generic operators with others of the same dimension has been obtained
with four distinct flavours, it is straightforward to apply it to
the appropriate operators of physical flavours.

We define the generic four fermion operators
\begin{equation}
\begin{array}{lcl}
O_{\Gamma^{(1)}\Gamma^{(2)}} &=& 
(\bar\psi_1\Gamma^{(1)}\psi_2)(\bar\psi_3\Gamma^{(2)}\psi_4)
\nn \\
O_{t^a \Gamma^{(1)} t^a \Gamma^{(2)}} &=& 
(\bar\psi_1\Gamma^{(1)}t^a \psi_2)(\bar\psi_3\Gamma^{(2)} t^a \psi_4)
\nn \\
O^F_{\Gamma^{(1)}\Gamma^{(2)}} &=& 
(\bar\psi_1\Gamma^{(1)}\psi_4)(\bar\psi_3\Gamma^{(2)}\psi_2)
\nn \\
O^F_{t^a \Gamma^{(1)} t^a \Gamma^{(2)}} &=& 
(\bar\psi_1\Gamma^{(1)}t^a \psi_4)(\bar\psi_3\Gamma^{(2)} t^a \psi_2)
\end{array}
\label{eq:qgam1gam1}
\end{equation}
where
$\Gamma^{(1)}$ and $\Gamma^{(2)}$ denotes any Dirac matrix, and $t^a$ the
colour matrices.
\begin{table}
\centering
\begin{tabular}{|r|r|r|r|r|r|}
\hline
$O_{\Gamma^{(1)} \Gamma^{(2)}}$ & $\ct{P}$ & $\ct{CS}'$ & $\ct{CS}''$ &
$\ct{CPS}'$ & $\ct{CPS}''$ \\ \hline \hline
$O_{VV}$ & $+1$ & $+1$ & $+1$ & $+1$ & $+1$ \\
$O_{AA}$ & $+1$ & $+1$ & $+1$ & $+1$ & $+1$ \\
$O_{PP}$ & $+1$ & $+1$ & $+1$ & $+1$ & $+1$ \\
$O_{SS}$ & $+1$ & $+1$ & $+1$ & $+1$ & $+1$ \\
$O_{TT}$ & $+1$ & $+1$ & $+1$ & $+1$ & $+1$ \\
\hline
$O_{[VA+AV]}$ & $-1$ & $-1$ & $-1$ & $+1$ & $+1$ \\
$O_{[VA-AV]}$ & $-1$ & $-1$ & $+1$ & $+1$ & $-1$ \\
$O_{[SP+PS]}$ & $-1$ & $+1$ & $+1$ & $-1$ & $-1$ \\
$O_{[SP-PS]}$ & $-1$ & $+1$ & $-1$ & $-1$ & $+1$ \\
$O_{T \tilde T}$ & $-1$ & $+1$ & $+1$  & $-1$ & $-1$  \\
\hline
\end{tabular}
\caption{Classification of four-fermion operators
according to lattice symmetries. These propeties are
also valid for the operators $O_{t^a \Gamma^{(1)} t^a \Gamma^{(2)}}$.
For the operators $O^F_{\Gamma^{(1)} \Gamma^{(2)}}$
and $O^F_{t^a \Gamma^{(1)} t^a \Gamma^{(2)}}$, we must
exchange the entries of the columns $\ct{CS}' \leftrightarrow \ct{CS}''$
and $\ct{CPS}' \leftrightarrow \ct{CPS}''$. }
\label{tab:g12}
\end{table}
Under renormalization, the operators of eq.(\ref{eq:qgam1gam1}) 
can in principle mix with any other dimension-six operator, provided it has
the same quantum numbers.
The generic QCD Wilson lattice action with 4 degenerate quarks
is symmetric under parity $\ct{P}$, and charge conjugation $\ct{C}$.
Moreover, there are three other useful (flavour) symmetries of the action,
namely the flavour exchange symmetry $\ct{S} \equiv (\psi_2
\leftrightarrow \psi_4)$ and the switching symmetries $\ct{S}' \equiv
(\psi_1 \leftrightarrow \psi_2 , \psi_3 \leftrightarrow \psi_4)$ and
$\ct{S}'' \equiv
(\psi_1 \leftrightarrow \psi_4 , \psi_2 \leftrightarrow \psi_3)$
\cite{Bernard}. In Table \ref{tab:g12} we classify the operators
$O_{\Gamma^{(1)} \Gamma^{(2)}}$ or combinations of them, 
according to the discrete symmetries
$\ct{P}$, $\ct{C}$, $\ct{S}'$ and $\ct{S}''$.  We adopt the notation
\[
O_{[\Gamma^{(1)}\Gamma^{(2)} \pm \Gamma^{(2)}\Gamma^{(1)}]} = 
O_{\Gamma^{(1)}\Gamma^{(2)}}\pm O_{\Gamma^{(2)}\Gamma^{(1)}}
\]
Note that the results of Table \ref{tab:g12} apply also to the
operators $O_{t^a \Gamma^{(1)} t^a \Gamma^{(2)}}$ since,
upon performing the symmetry transformations, sign differences,
resulting from the presence of the colour $t^a$ matrix, disappear because
the colour matrices appear quadratically.
On the other hand, $O^F_{\Gamma^{(1)} \Gamma^{(2)}}$ is
obtained by applying $\ct{S}$ on $O_{\Gamma^{(1)} \Gamma^{(2)}}$.
Since $\ct{S}$ transforms $\ct{S}'$ into $\ct{S}''$, the properties
of Table \ref{tab:g12} also apply to
$O^F_{\Gamma^{(1)} \Gamma^{(2)}}$, but with all $\ct{S}'$
and $\ct{S}''$ columns exchanged. Again, the operator
$O^F_{t^a \Gamma^{(1)} t^a \Gamma^{(2)}}$ has the same properties
as $O^F_{\Gamma^{(1)} \Gamma^{(2)}}$, since the colour matrix
$t^a$ appears quadratically.

Our aim is to find complete bases of operators which mix under
renormalization. Thus, besides classifying them according to their
symmetries, we must also eliminate the operators which are not
independent.  This is seen by
applying the standard identity of colour matrices
\begin{equation}
t^a_{AB} t^a_{CD} = -\frac{1}{2N} \delta_{AB} \delta_{CD}
+\frac{1}{2} \delta_{AD} \delta_{CB}
\label{eq:tt}
\end{equation}
on the $t^a$'s of a given operator. 
For the operator $O_{t^a \Gamma^{(i)} t^a \Gamma^{(j)}}$
the result has the general form
\[
O_{t^a \Gamma^{(i)} t^a \Gamma^{(j)}} = - \frac{1}{2N_c}
O_{\Gamma^{(i)} \Gamma^{(j)}} + \frac{1}{2} \sum_{n,m}
C_{nm} O^F_{\Gamma^{(n)} \Gamma^{(m)}}
\]
where the sum runs over all the Dirac matrices obtained by the
Fierz transformation of $\Gamma^{(i)} \Gamma^{(j)}$, and the factors
$C_{nm}$ are the appropriate constants of the Fierz transformation 
\cite{DI=3/2}.
Analogously we can express $O^F_{t^a \Gamma^{(i)} t^a \Gamma^{(j)}}$
in terms of $O_{\Gamma^{(i)} \Gamma^{(j)}}$ and
$O^F_{\Gamma^{(i)} \Gamma^{(j)}}$.
Therefore, in the following,  it is adequate to limit ourselves to the
mixing of $O_{\Gamma^{(i)} \Gamma^{(j)}}$'s and
$O^F_{\Gamma^{(i)} \Gamma^{(j)}}$'s, according to the entries of Table
\ref{tab:g12}. 

Having eliminated the non-independent operators
we proceed in classifying the complete bases of operators which mix
under renormalization, according to the following two rules:
\begin{enumerate}
\item
All operators with identical values of $\ct{P}$,
$\ct{CPS}'$ and $\ct{CPS}''$ are allowed to mix with each other. These
now form a maximal basis.
\item
If possible, the maximal basis must be decomposed into smaller bases, by 
using  the remaining symmetry $\ct{S}$, in order to form linear
combinations of the operators of our basis, which have definite $\ct{S}$, 
i.e.\ $\ct{S}=+1$ or $\ct{S}=-1$, that can only mix among themselves. 
\end{enumerate}

The first rule is easy to apply in practice, because 
the $\ct{P}$, $\ct{CPS}'$ and $\ct{CPS}''$ values of the operators can be
read-off from Table \ref{tab:g12} . 
As an example of this rule, we note that
$O_{[SP - PS]}$ mixes with with $O^F_{[VA - AV]}$, since they both have
$\ct{P} = -1$, $\ct{CPS}' = -1$ and $\ct{CPS}''=+1$.
Having thus applied the first rule, we turn to the specific task of
reducing the basis, for each case of interest (second rule).
This we now do case by case, using $\ct{S}$ symmetry.

\subsection{Parity violating operators}

We consider first the parity violating four-fermion operators, cf.
Tab.~\ref{tab:g12}.
None of the four violating operators
have identical
$\ct{CPS}'$ and $\ct{CPS}''$ values. 
Each of them, however, mixes with some $\ct{S}$-counterpart, e.g.
$O_{[SP + PS]}$ and $O^F_{[SP + PS]}$ or $O_{[VA - AV]}$, $O^F_{[SP - PS]}$.

We examine first $O_{[VA + AV]}$ which  mixes with
$O^F_{[VA + AV]}$ only, forming a basis 
of two operators, characterized by $\ct{CPS}'= \ct{CPS}'' = +1$.
We rotate this basis into
\be
O^\pm_{[VA+AV]} \equiv \frac{1}{2} \left [  O_{[VA + AV]}  \pm
O^F_{[VA + AV]} \right ],
\label{vaav24}
\ee
and note that $O^+_{[VA+AV]}$ has $\ct{S} = +1$
and $O^-_{[VA+AV]}$ has $\ct{S} = -1$. Thus, they do not mix with
each other. The final result is that the original basis of two
operators has been decomposed into two bases of one operator each:
the two operators $O^\pm_{[VA+AV]}$ of eq.(\ref{vaav24}) renormalize
multiplicatively.

We now turn to $O_{[VA - AV]}$. It mixes with
$O^F_{[SP - PS]}$, since they both have
$\ct{CPS}' = +1$ and $\ct{CPS}'' = -1$. Similarly, 
$O^F_{[VA - AV]}$ and $O_{[SP - PS]}$ have
$\ct{CPS}' = -1$, $\ct{CPS}'' = +1$.
It is convenient to combine the two bases into a product basis
of 4 operators:
\begin{equation}
\begin{array}{l}
O^\pm_{[VA-AV]} \equiv \frac{1}{2} \left [(O_{[VA - AV]}
 \pm  O^F_{[VA - AV]} \right ] \\
O^\pm_{[SP-PS]} \equiv \frac{1}{2} \left [ O_{[SP - PS]} \pm
O^F_{[SP - PS]} \right ]
\end{array}
\label{sppsm24}
\end{equation}
None of these operators have definite $\ct{CPS}'$ or $\ct{CPS}''$.
However, they have definite $\ct{S} = \pm 1$. 
Thus, they mix in pairs according to their $\ct{S}$ value;
i.e. we have reduced the original basis of four operators into two
bases of two operators each.

Similarly, we rotate the operators
$O_{[SP + PS]}$, $O^F_{[SP + PS]}$, $O_{T \tilde T}$ and $O^F_{T \tilde T}$
(with $\ct{CPS}'=\ct{CPS}'' = -1$), into the new basis 
\begin{equation}
\begin{array}{l}
O^\pm_{[SP+PS]} \equiv \frac{1}{2} \left [  O_{[SP + PS]}  \pm
O^F_{[SP + PS]} \right ] \\
O^\pm_{T \tilde T} \equiv \frac{1}{2} \left [  O_{T \tilde T}  \pm
O^F_{T \tilde T} \right ]
\end{array}
\label{spps24}
\end{equation}
which, once more, is decomposed into two bases, of two operators each, with
definite $\ct{S} = \pm 1$.

If we introduce, for notational compactness, the notation
\begin{equation}
\begin{array}{l}
\bar Q^\pm_1 \equiv O^\pm_{[VA+AV]} \\
\bar Q^\pm_2 \equiv O^\pm_{[VA-AV]} \\
\bar Q^\pm_3 \equiv O^\pm_{[SP-PS]} \\
\bar Q^\pm_4 \equiv O^\pm_{[SP+PS]} \\
\bar Q^\pm_5 \equiv O^\pm_{T \tilde T} 
\end{array}
\label{eq:PVbasis}
\end{equation}
the renormalization structure becomes
\be
\label{eq:mrenpv}
\hat {\bar Q} _i^\pm = \bar \ctZ^\pm_{ij} 
\bar Q_j^\pm \qquad (i,j = 1,\dots ,5)
\ee
where $\hat{\bar Q}_i^\pm$ denote the renormalized
operators and $\bar \ctZ_{ij}^\pm$ is the
renormalization (and mixing) matrix (summation over repeated indices is
implied). 

Dropping, for simplicity, the $\pm$ subscripts, 
the matrix $\bar \ctZ_{ij}$ is a (relatively sparse) block diagonal matrix
of the form
\begin{equation}
\left(\begin{array}{rrrrr}
\bar \ctZ_{11} & 0 & 0 & 0 & 0 \\
0 & \bar \ctZ_{22} & \bar \ctZ_{23} & 0 & 0 \\
0 & \bar \ctZ_{32} & \bar \ctZ_{33} & 0 & 0 \\
0 & 0 & 0 & \bar \ctZ_{44} & \bar \ctZ_{45} \\
0 & 0 & 0 & \bar \ctZ_{54} & \bar \ctZ_{55}
\end{array}\right)
\label{eq:renpc}
\end{equation}
It is important to notice that this NP renormalization structure, determined
by the symmetries of the action, is the same as in the continuum na\"\i ve
dimensional regularization scheme, or any other regularization that does not
break chirality explicitly.  In fact, the operators that mix belong to the
same chiral representation, and their chiral structures can be obtained from
each other by Fierz transformations.

\subsection{Parity conserving operators}
\label{subsec:pcbases}

Let us now pass to the parity-conserving operators, cf.\ Tab. 1.
All of the parity conserving operators $O_{\Gamma \Gamma}$ 
are eigenstates of all the discrete symmetries listed, with
eigenvalue $+1$.  
Thus by rule 1., unlike the parity violating case, they all mix among each 
other and also with the five $O^F_{\Gamma \Gamma}$'s; the complete
maximal basis consists of 10 operators.
By rule 2., we rotate our basis into a new one:
\begin{equation}
O^\pm_{\Gamma \Gamma} = \frac{1}{2} \left [
O_{\Gamma \Gamma} \pm O^F_{\Gamma \Gamma} \right ] 
\label{eq:qgamgam}
\end{equation}
in which the 5 $O^+_{\Gamma \Gamma}$'s have being $\ct{S}=+1$ and mix
only among themselves; the same is true for the 
$O^-_{\Gamma\Gamma}$'s which have $\ct{S}=-1$. Thus the original 
basis of 10 operators has been decomposed into two independent
bases of 5 operators each.
\par
This result can be used in the renormalization of the operators
$O^\pm_{[VV + AA]}$, which are the 
parity conserving parts of the operators $O^\pm_{LL}$.
Clearly, $O^\pm_{[VV + AA]}$ mixes with
$O^\pm_{[VV - AA]}$, $O^\pm_{SS}$, $O^\pm_{PP}$
and $O^\pm_{TT}$. Any other linear combination of these mixing operators
of ``wrong'' naive chirality 
is in principle acceptable; particular choices are a question of convenience.
Here we discuss three such options.

The first option is the one which enables a comparison
of this mixing to the perturbative calculations
of \cite{PT_4f} (and also the NP computations of \cite{DS=2,B_K,DI=1/2}). We 
call this a {\em perturbative-inspired} (PI) basis:
\begin{equation}
\begin{array}{l}
O^{\pm}_0\equiv\frac{1}{4}O^{\pm}_{[VV+AA]} \\
O^{\pm}_1\equiv-\frac{1}{8N_c}O^{\pm}_{[SS-PP]} \\
O^{\pm}_2\equiv-\frac{(N_c^2\pm N_c-1)}{16N_c}O^{\pm}_{[VV-AA]} \\
O^{\pm}_3\equiv\frac{(\pm N_c-1)}{8N_c}O^{\pm}_{[SS+PP+TT]} \\
O^{\pm}_4\equiv\frac{(\pm N_c-1)}{8N_c}O^{\pm}_{[SS+PP-\frac{1}{3}TT]} 
\end{array}
\label{eq:PIbasis}
\end{equation}
In this base, $O^\pm_0$ mixes with the operators $O^{\pm}_i,\ i=1,2,3$ 
which already appear at the level of the one-loop perturbative calculation 
\cite{PT_4f}, but also with $O^{\pm}_4$ which is not present at the one-loop 
level.  The arbitrary numerical overall colour
factors of $O^{\pm}_i,\ i=1,2,3$ are 
defined so as to be in agreement with the convention of 
\cite{PT_4f}, and the colour factor of $O^{\pm}_4$ has been set equal to the 
one of $O^{\pm}_3$. This choice seems natural for the comparison of
its relative weight with respect to the other  operators present at one-loop.

A second option, which is exploited in \cite{JLQCD}, consists in taking
the basis of the eigenvector of the Fierz matrix.  We then call this 
a {\em Fierz-inspired} (FI) basis:
\begin{equation}
\begin{array}{l}
{\cal O}^\pm_1 \equiv O^\pm_{[VV+AA]} \nonumber \\
{\cal O}^\pm_2 \equiv O^\pm_{[SS+PP+TT]} \nonumber \\
{\cal O}^\pm_3 \equiv O^\pm_{[SS+PP-\frac{1}{3}TT]} \\
{\cal O}^\pm_4 \equiv O^\pm_{[VV-AA+2(SS-PP)]} \nonumber \\
{\cal O}^\pm_5 \equiv O^\pm_{[VV-AA-2(SS-PP)]} \nonumber
\end{array}
\label{eq:FIbasis}
\end{equation}

A third option, used in \cite{DI=3/2}, consists in the following basis, 
which we call {\em Ward-inspired} basis, as it can obtained, up to signs, from 
the parity-violating base, eq.~(\ref{eq:PVbasis}), with a
chiral transformation $\psi_4\to \gamma_5\psi_4$:
\begin{equation}
\begin{array}{l}
Q^\pm_1 \equiv O^\pm_{[VV+AA]} \nonumber \\
Q^\pm_2 \equiv O^\pm_{[VV-AA]} \nonumber \\
Q^\pm_3 \equiv O^\pm_{[SS-PP]} \\
Q^\pm_4 \equiv O^\pm_{[SS+PP]} \nonumber \\
Q^\pm_5 \equiv O^\pm_{TT} \nonumber
\end{array}
\label{eq:WIbasis}
\end{equation}
Whichever basis we choose, upon renormalization the structure will be 
of the form
\be
\label{eq:mrenpc}
\hat Q_i^\pm = \ctZ^\pm_{ij} Q_{j}^\pm \qquad (i,j = 1,\dots ,5),
\ee
where the matrix $\ctZ^\pm$ is not sparse as in the PV case.

\subsection{Scale dependence}

Close to the continuum and chiral limit,
the UV divergent elements of the renormalization matrix $\ctZ$
depend on $a\mu$ and $g_0^2$ only, $\ctZ_{ij} = \ctZ_{ij}(a\mu,g_0^2)$,
whereas non-divergent elements are of the form $\ctZ_{ij}(g_0^2)$.
Since the structure of the renormalization matrix $\bar \ctZ$ is the same
as in the continuum, all
its matrix elements are logarithmically divergent, i.e. $\bar \ctZ_{ij}
= \bar \ctZ_{ij}(a\mu,g_0^2)$.

The specification of which elements of the matrix $\ctZ$ diverge and
which are finite, in the limit $a \rightarrow 0$, 
can be achieved non-perturbatively with the aid of the axial Ward
Identity (WI).  We refer to \cite{DI=3/2} for a detailed presentation while
we give here only the prescription for the renormalization.  
First, we need to construct the subtracted 
operators
\begin{equation}
\begin{array}{l}
Q_1^{\sub}=Q_1+\sum_{j=2}^5 c_{1j} Q_{j}, \\
Q_{i}^{\sub}=Q_{i}+\sum_{j=1,4,5}c_{ij} Q_{j},\ \ i=2,3 \\
Q_{i}^{\sub}=Q_{i}+\sum_{j=1,2,3}c_{ij} Q_{j},\ \ i=4,5 
\end{array}
\end{equation}
and, second, renormalize
the operators $Q_i^{\sub}$ as in the continuum, i.e.\ with a matrix of the
form given in the PV case, cf. eq.~(\ref{eq:renpc}).

Before discussing the numerical results, it is worth while to 
stress in which conditions the direct implementation of the WI, 
as exploited for the $\Delta S=2$ operator in \cite{JLQCD},
and the NPM are equivalent.
The WI holds for operators with the correct chiral properties, that is
multiplicatively renormalizable operators transforming according to a well
defined representation of the chiral algebra. In fact, it is by imposing
its validity on the renormalized operators one can fix the mixing coefficients
of the form factors which stem from the chiral violation due to the Wilson 
term.  In the NPM, this is achieved by 
imposing that the projections of the renormalized operator 
$\widehat O^{\Delta S=2}$ 
(cf.~eq.~(\ref{eq:O_DS=2}) for the its explicit form) 
on the four chiral violating form factors are zero,
\begin{equation}
\Pr_j\<p|\widehat O^{\Delta S=2}|p\>=0,\ j=1,\ldots,4.
\end{equation}
But this is true if there are no other causes of chiral symmetry breaking, 
either due to explicit presence of mass terms or due to 
spontaneous symmetry breaking in the chiral limit.  
But both these effect die off in the large momenta region.
So the WI method and the NPM are equivalent for sufficiently large values of 
the renormalization scale $\mu^2=p^2$.
In any case, the overall multiplicatively renormalization constant cancels
in the WI and thus cannot be determined.  Thus the NPM (or some other 
renormalization method) is needed even if the WI method is used to obtain
the mixing coefficients.

\section{Numerical results}

% figure ZplusmixPIk1432
\begin{figure}[t]
\centerline{\hspace{2cm}\epsfig{figure=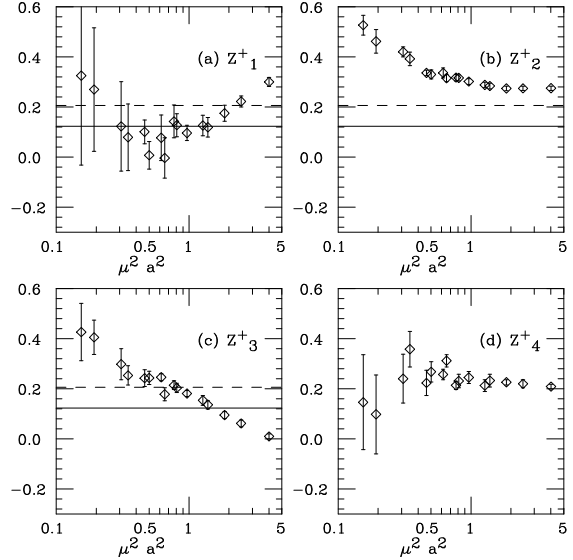,height=10cm,angle=90}}
\caption{NP $\Delta S=2$ mixing $Z^+$'s in PI basis as a function of $\mu^2a^2$ for $\kappa=0.1432$.  The solid (dashed) line is from SPT (BPT).}
\label{fig:ZplusmixPIk1432}
\end{figure}

% figure ZplusmixPI
\begin{figure}[t]
\centerline{\hspace{2cm}\epsfig{figure=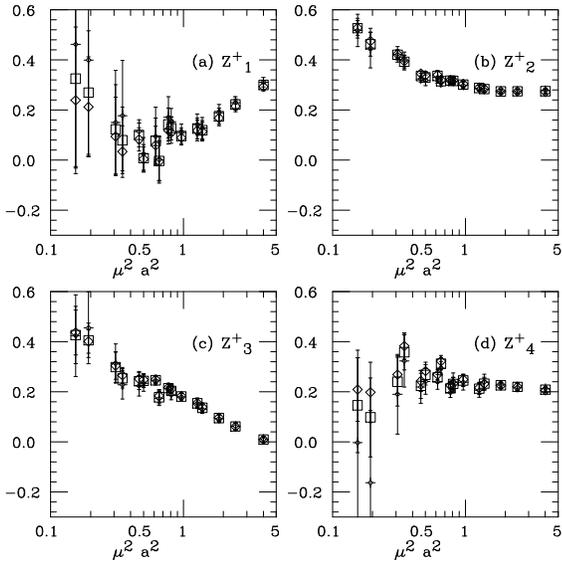,height=10cm,angle=90}}
\caption{$m_qa$ dipendence for NP $\Delta S=2$ mixing $Z^+$'s in PI basis as a function of $\mu^2a^2$.  The three symbols correspond to the values of 
$\kappa=0.1425,0.1432,0.1440$.}
\label{fig:ZplusmixPI}
\end{figure}

\begin{table}[t]
\begin{tabular}{|r|r|r|r|}
\hline
$\mu^2 a^2$ & $\alpha$ & $\beta$ & $\gamma$ \\ \hline \hline
0.31 & $ 0.030(18)$ &  0.27(21) & 0.90(15) \\ \hline
0.62 & $-0.027(16)$ &  0.36(18) & 0.75(13) \\ \hline
0.96 & $-0.012(14)$ &  0.24(17) & 0.69(12) \\ \hline
1.27 & $ 0.005(13)$ &  0.14(16) & 0.68(12) \\ \hline
1.39 & $-0.009(13)$ &  0.24(16) & 0.67(12) \\ \hline
1.85 & $-0.003(13)$ &  0.18(16) & 0.66(11) \\ \hline
2.46 & $-0.001(12)$ &  0.24(15) & 0.65(11) \\ \hline
4.01 & $-0.002(12)$ &  0.44(15) & 0.67(11) \\ \hline
 BPT & $-0.052(12)$ &  0.16(15) & 0.62(11) \\ \hline
\end{tabular}
\caption{Values of $\alpha,\beta,\gamma$ for different values of $\mu a$.}
\label{tab:params}
\end{table}

% figure chiralplus
\begin{figure}[t]   
\centerline{\epsfig{figure=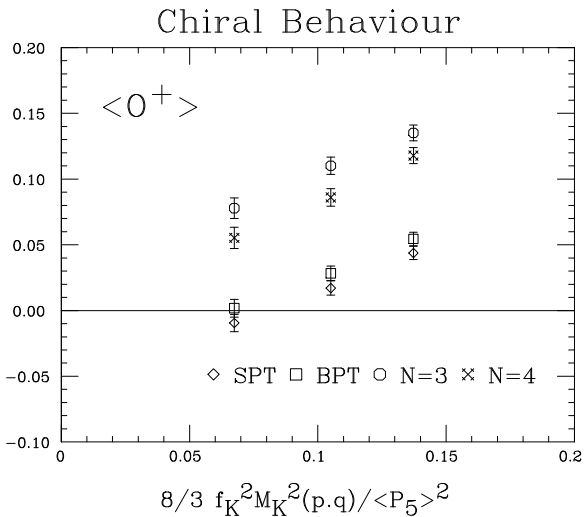,height=10cm,angle=90}}
\caption{Chiral behaviour of $B_K$.  The NP $Z$'s are taken at $\mu^2a^2=0.96$.}
\label{fig:chiralplus}
\end{figure}

% figure ZplusmixFIk1432
\begin{figure}[t]   
\centerline{\hspace{2cm}\epsfig{figure=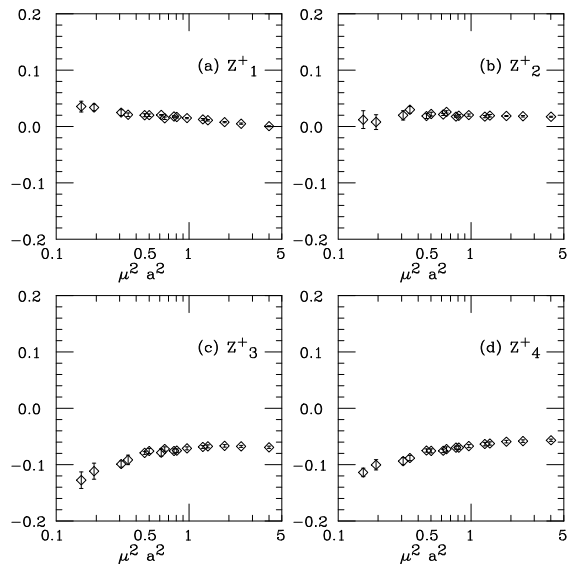,height=10cm,angle=90}}
\caption{NP $\Delta S=2$ mixing $Z^+$'s in FI basis as a function of $\mu^2a^2$ for $\kappa=0.1432$.}
\label{fig:ZplusmixFIk1432}
\end{figure}

% figure Zplusk1432
\begin{figure}[t]
\centerline{\epsfig{figure=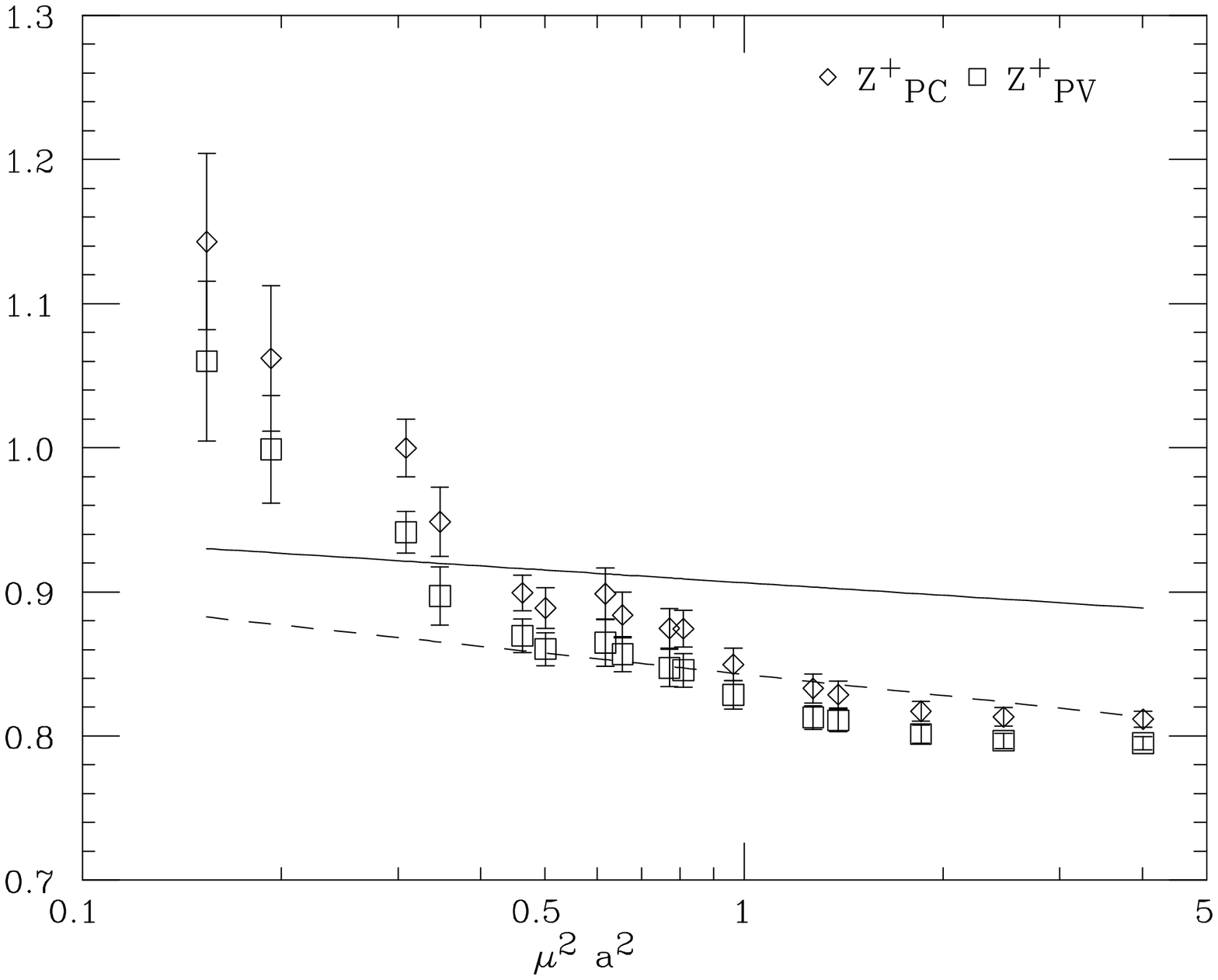,height=10cm,angle=90}}
\caption{Overall NP LL $Z^+$ as a function of $\mu^2a^2$.  
The solid (dashed) line is from SPT (BPT).}
\label{fig:Zplusk1432}
\end{figure}

% figure Zminusk1432
\begin{figure}[t]
\centerline{\epsfig{figure=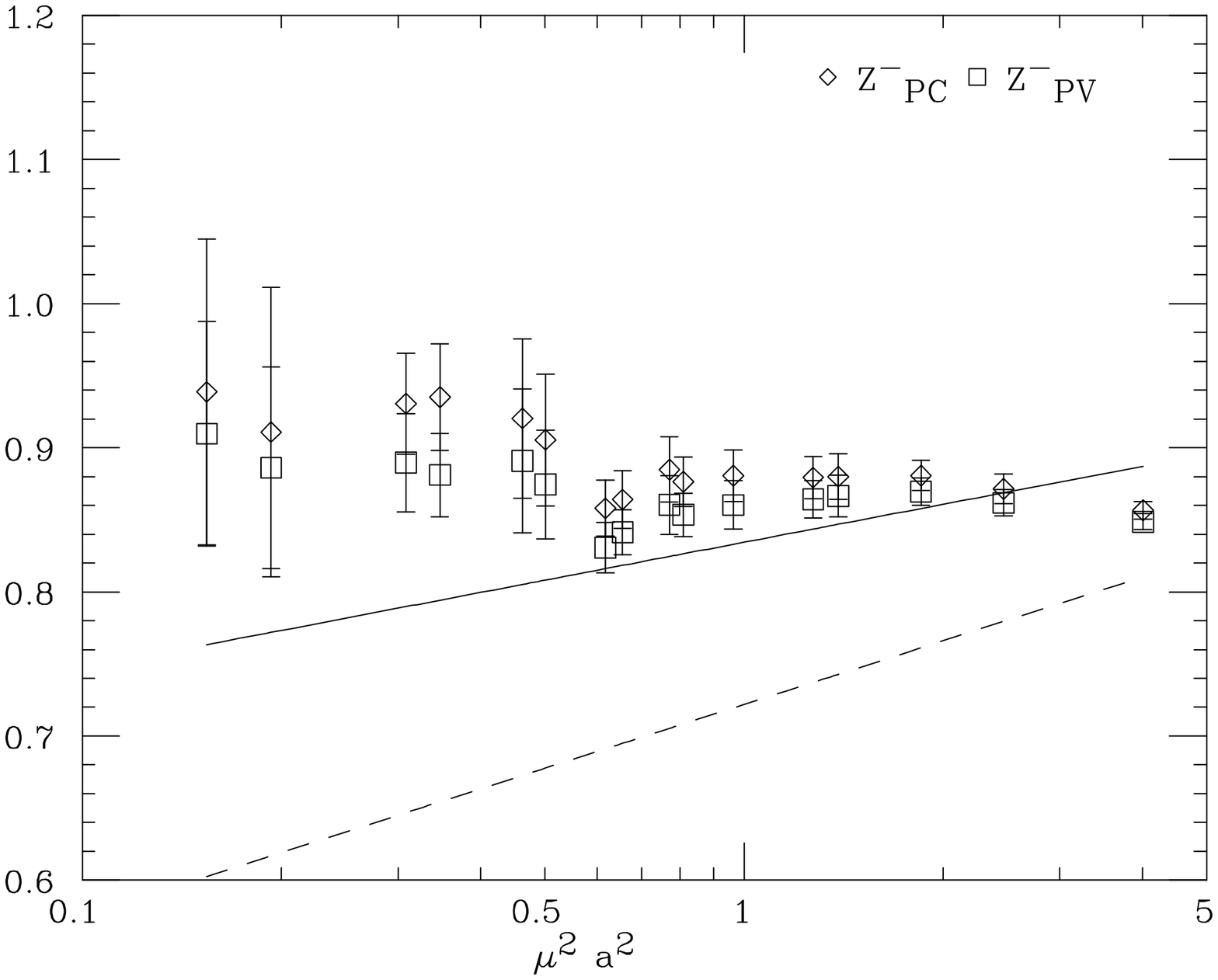,height=10cm,angle=90}}
\caption{Overall NP LL $Z^-$ as a function of $\mu^2a^2$.  
The solid (dashed) line is from SPT (BPT).}
\label{fig:Zminusk1432}
\end{figure}

Our NP Monte Carlo simulations have been performed on an APE machine.
We have generated an ensemble of 100 independent gauge-field configurations,
using a tree-level improved SW-Clover action \cite{SW,Heatlie} action
on a $16^3\times 32$ lattice, at $\beta=6.0$.  The quark propagator 
has been calculated, in the lattice Landau gauge,
for three values of the hopping parameter $\kappa=0.1425,0.1432,0.1440$,
corresponding to a pion mass of $900-600$ MeV.  
In comparing the NP results with PT, we have used both 
a standard bare lattice coupling (SPT) $\alpha_s^{\latt}=g_0^2/4\pi$ and 
boosted coupling (BPT) of 
$\alpha_V=\alpha_s^{\latt}/\<\Tr \Box\>\simeq\, 1.68\,\alpha_s^{\latt}$ 
\cite{Lepage}.

The three values of the hopping parameter allow us to extend the analysis 
presented in \cite{DI=1/2} to include the study of the mass dependence 
of the $Z$'s, i.e. the effects of the systematics of $O(m_qa)$. 
In general, these effects were expected to be small for the light quark
sector \cite{m_qa}, though at lower values of $m_qa$ the statistical errors
tend to be larger and at higher values of $m_qa$ the discretization errors
are larger.  We have found that indeed the expectations are fulfilled, as
can be seen from fig.~\ref{fig:ZplusmixPI} in which the symbols representing
the different $\kappa$ values can barely be distinguished, at least in the
significant region of large $\mu^2a^2$.  Thus, we 
have chosen to present the results at the intermediate $\kappa=0.1432$.
We have chosen not to extrapolate in $m_qa$ as we feel that the best one can
do with a systematic error is to control it rather than extrapolate in it. 
 
We first consider the renormalization of the 
operator $O^+_0$ of eq.~(\ref{eq:PIbasis}), which has the same renormalization 
properties of parity-conserving part of the $\Delta S=2$ operator 
$\bar s \gamma^L_{\mu} d\bar s \gamma^L_{\mu} d$.  The renormalized operator
is 
\begin{equation}
\hat O^+=Z^+_0(\mu a)[O^+_0(a)+\sum_{i=1}^4 Z^+_i O^+_i(a)],
\label{eq:O_DS=2}
\end{equation}
where the operators $O^+_1,\ldots,O^+_4$ are given in eq.~(\ref{eq:PIbasis}) 
for the PI basis, and by ${\cal O}^+_2,\ldots,{\cal O}^+_5$ in 
eq.~(\ref{eq:FIbasis}) for the FI basis.  

In fig.~\ref{fig:ZplusmixPIk1432} we show the results of the mixing $Z$'s 
in the PI basis at different renormalization scales $\mu^2a^2$. 
It is clearly notable that that $Z^+_2$ and $Z^+_4$ are very well defined and 
almost scale-independent in a large ``window'' of $\mu^2a^2$, 
whereas $Z^+_1$ and $Z^+_3$
present a smaller window, i.e.\ a more pronounced scale-dependence.
Moreover, $Z^+_4$ which is absent in 1-loop PT, is not negligible.
We stress that the large fluctuations at small $\mu^2a^2$ do not 
spoil the validity of the NPM, since in that region the perturbative
matching to a continuum scheme is not reliable, as for any
NP lattice method at a fixed lattice spacing \cite{Martinelli}.
Fig.~\ref{fig:ZplusmixPI} shows the $m_qa$-dependence of the $Z$'s in the PI
basis.  As anticipated, the systematic error of $O(m_qa)$ is very small,
at least in the region of significant values of $\mu^2a^2$,
for the quark masses we used. 

Using the NP $Z^+$'s in PI basis with their mass dependence taken into 
account (extending the analysis of ref.~\cite{DI=1/2}),
we can revisite the study of the chiral 
behaviour of the $B_K$ parameter \cite{DI=3/2,B_K}.
We adopt the usual parametrization
\begin{equation}
\<O^+\>=\alpha+\beta m_K^2+\gamma(p\cdot q)+...
\end{equation}
where $\alpha$ is a lattice artefact that we expect to vanish in the chiral 
limit. In tab.~\ref{tab:params} we present the parameters $\alpha,\beta,\gamma$
obtained for different scales $\mu^2a^2$.   
In fig.\ \ref{fig:chiralplus} the result using the NP $Z^+$'s at 
$\mu^2=0.96$ and the bare matrix elements from \cite{B_K} is shown. 
Clearly, the use of the mixing $Z$'s in PT (all equal at 1-loop \cite{PT_4f}) 
does not yield the desired behaviour, even if a boosted coupling is used.
This is due to the delicate cancellations which occur among the bare
matrix elements that can only be resolved beyond 1-loop.
On the contrary, using the NP $Z$'s, the intercept is compatible with
zero.  This behaviour is consistently found at all scale 
$\mu^2a^2\gtapprox 0.96$.  
Although the use of the complete set of operators ($N=4$ in 
fig.\ \ref{fig:chiralplus})
yield a better chiral behaviour that the use of only the operators which
mix in 1-loop PT $(N=3)$, since the mixing with $O^+_4$ starts at 
${\cal O}(g_0^4)$, we do not expect 
drastic changes in the chiral behaviour.  If the chiral behaviour were
sensibly different, we would not trust the matching to the continuum which
has an uneludable perturbative uncertainty.  Indeed we find that $\alpha^{N=3}$
is also compatible with zero, and compatible with $\alpha^{N=4}$ although 
with large statistical errors forced by the thinning
approximation \cite{B_K}.   We can only state that 
the value of $B_K$, proportional to $\gamma$, is unaltered and its RGI value 
at NLO is $\hat B_K=0.85\pm 0.15$ while the correct chiral behaviour of the 
continuum, signaled by the vanishing of $\alpha$, is recovered \cite{DI=3/2}.  

The chiral behaviour of the $\Delta S=2$ has also been studied in \cite{JLQCD}
imposing the chiral WI on quark states using the FI basis.  
The results for the mixing $Z$'s, obtained with an unimproved
Wilson action, seems to show a very stable signal as a function of $\mu^2a^2$
\cite{Kuramashi}. 
In particular, the fluctuations at small scales are much reduced.
It could be concluded that the WI method and the NPM, although conceptually
equivalent at large values of $\mu$, are not numerically such.  
To understand this point in more detail, we have tried taking linear 
combinations of the $Z$'s in PI basis and expressing them in the FI basis, 
as shown in fig.~\ref{fig:ZplusmixFIk1432}.
It is clear that the $Z$'s, calculated with the NPM, in this new basis show 
a much greater stability, comparable to the one obtained by the WI method.
It must be stressed that although the stability of the $Z$'s depends on the
choice of the basis,  the physical results do not.  In fact, the correct
chiral behaviour of the renormalized operator is obtained with either basis.
This is due to the fact that the $Z$'s which present greater fluctuations
multiply bare matrix elements of the operators that weight less than
the ones multiplied by the more stable $Z$'s.  So that stability of the
chiral behaviour with the PI basis shown in tab.~\ref{tab:params} 
is due to the extremely clean determination of $Z_2$ in 
fig.~\ref{fig:ZplusmixPIk1432} and to the fact that 
$\<O_2\>\approx -3\<O_{1,3,4}\>$ \cite{DI=3/2}.

As a final flourish, in fig.~\ref{fig:Zplusk1432} and 
fig.~\ref{fig:Zminusk1432} we show the overall renormalization constants for
$O^{\pm}_{[VV+AA]}$ and $O^{\pm}_{[VV-AA]}$, which are the PC and PV part
of the LL operator and, as already stressed, can only be determined with 
the NPM.  The comparison with PT shows that, while $Z^+$ is in good agreement
with boosted PT, for $Z^-$ the agreement is better with standard PT.

\section{Conclusions}

Recently, there has been considerable progress both in the Symanzik 
improvement program and in the development of non-perturbative renormalization
methods.  We have presented the application of the off-shell renormalization
method using a tree-level improved SW-Clover action to four-fermion operators
with light quarks relevant to weak decays.  For the $\Delta I=3/2$ sector we
have presented a general analysis of the renormalization structure, and applied
it to the $\Delta S=2$ operator obtaining the correct chiral behaviour for
the $B_K$ parameter.  On the other hand, for the $\Delta I=1/2$ sector,
the standard positive parity approach was shown to be unviable, and
new ideas are needed.

\section*{Acknowledgements}

I am grateful to A. Donini, V. Gim\`enez, G. Martinelli, G.C. Rossi, 
C.T. Sachrajda, S. Sharpe, M. Testa and A. Vladikas, 
for a most enjoyable and fruitful collaboration on the material presented 
in this talk.  I would like to thank A. Vladikas for reading the manuscript
and for his comments.

I would also like to extend a warm thank to the organizers of the Workshop 
and to the Center for Computational Physics in Tsukuba, 
for their invitation and for creating a very stimulating and pleasant 
scientific environment.

I acknowledge support from EPSRC through grant GR/K41663, from PPARC 
through grant GR/L22744 and partial support from INFN.

\end{document}